\def\papertitle{Autoencoding Neural Networks as Musical Audio Synthesizers}
\def\paperauthorA{Joseph Colonel}
\def\paperauthorB{Christopher Curro}
\def\paperauthorC{Sam Keene}
\def\paperauthorD{Author Four}

\documentclass[twoside,a4paper]{article}
\usepackage{dafx_18}
\usepackage{amsmath,amssymb,amsfonts,amsthm}
\usepackage{euscript}
\usepackage[latin1]{inputenc}
\usepackage[T1]{fontenc}
\usepackage{ifpdf}

\usepackage[english]{babel}
\usepackage{caption}
\usepackage{subfig} % or can use subcaption package
\usepackage{color}

\usepackage[group-separator={,}]{siunitx}

\ninept

\usepackage{times}
\newif\ifpdf
\ifx\pdfoutput\relax
\else
   \ifcase\pdfoutput
      \pdffalse
   \else
      \pdftrue
\fi

\ifpdf % compiling with pdflatex
  \usepackage[pdftex,
    pdftitle={\papertitle},
    pdfauthor={\paperauthorA, \paperauthorB, \paperauthorC, \paperauthorD},
    colorlinks=false, % links are activated as colror boxes instead of color text
    bookmarksnumbered, % use section numbers with bookmarks
    pdfstartview=XYZ % start with zoom=100% instead of full screen; especially useful if working with a big screen :-)
  ]{hyperref}
  \usepackage[pdftex]{graphicx}
  \usepackage[figure,table]{hypcap}
\else % compiling with latex
  \usepackage[dvips]{epsfig,graphicx}
  \usepackage[dvips,
    colorlinks=false, % no color links
    bookmarksnumbered, % use section numbers with bookmarks
    pdfstartview=XYZ % start with zoom=100% instead of full screen
  ]{hyperref}
  \usepackage[figure,table]{hypcap}
\fi

\title{\papertitle}

\threeaffiliations{
\paperauthorA \,}
{\href{https://ee.cooper.edu/~colone}{The Cooper Union for the Advancement of Science and Art} \\ NYC, New York, USA\\
{\tt \href{mailto:colone@cooper.edu}{colone@cooper.edu}}
}
{\paperauthorB \,}
{\href{}{The Cooper Union for the Advancement of Science and Art} \\ NYC, New York, USA\\
{\tt \href{mailto:curro@cooper.edu}{curro@cooper.edu}}
}
{\paperauthorC \,}
{\href{https://ee.cooper.edu/~keene}{The Cooper Union for the Advancement of Science and Art} \\ NYC, New York, USA\\ {\tt \href{mailto:keene@cooper.edu}{keene@cooper.edu}}
}
\begin{document}
% more pdf-tex settings:
\ifpdf % used graphic file format for pdflatex
  \DeclareGraphicsExtensions{.png,.jpg,.pdf}
\else  % used graphic file format for latex
  \DeclareGraphicsExtensions{.eps}
\fi

\maketitle

\begin{abstract}
A method for musical audio synthesis using autoencoding neural networks is proposed. The autoencoder is trained to compress and reconstruct magnitude short-time Fourier transform frames. The autoencoder produces a spectrogram by activating its smallest hidden layer, and a phase response is calculated using real-time phase gradient heap integration. Taking an inverse short-time Fourier transform produces the audio signal. Our algorithm is light-weight when compared to current state-of-the-art audio-producing machine learning algorithms.  We outline our design process, produce metrics, and detail an open-source Python implementation of our model.
\end{abstract}

\section{Introduction}
\label{sec:intro}
There are many different methods of digital sound synthesis. Three traditional methods are additive, subtractive, and frequency modulation (FM) synthesis. In additive synthesis, waveforms such as sine, triangle, and sawtooth waves are generated and added to one another to create a sound. The parameters of each waveform in the sum are controlled by the musician. In subtractive synthesis, a waveform such as a square wave is filtered to subtract and alter harmonics. In this case, the parameters of the filter and input waveform are controlled by the musician. Lastly, in FM synthesis the timbre of a waveform is generated by one waveform modulating the frequency of another. In this method, musicians control the parameters of both waveforms, and the manner in which one modulates the other. 

Recently, machine learning techniques have been applied to musical audio sythesis. One version of Google's Wavenet architecture uses convolutional neural networks (CNNs) trained on piano performance recordings to prooduce raw audio one sample at a time \cite{deepmind}. The outputs of this neural network have been described as sounding like a professional piano player striking random notes. Another topology, presented by Dadabots, uses recurrent neural networks (RNNs) trained to reproduce a given piece of music \cite{dadabots}. These RNNs can be given a random initialization and then left to produce music in batches of raw audio samples. Another Google project, Magenta \cite{magenta}, uses neural network autoencoders (autoencoders) to interpolate audio between different instrument's timbres. While all notable in scope and ability, these models require immense computing power to train and thus strip musicians of full control over the tools. 

In this paper, we present a new method for sound synthesis that incorporates deep autoencoders while remaining light-weight. This method is based off techniques for constructing audio-handling autoencoders outlined in \cite{colonel}. We first train an autoencoder to encode and decode magnitude short-time Fourier transform (STFT) frames generated by audio recorded from a subtractive synthesizer. This training corpus consists of five-octave C Major scales on various synthesizer patches. Once training is complete, we bypass the encoder and directly activate the smallest hidden layer of the autoencoder. This activation produces a magnitude STFT frame at the output. Once several frames are produced, phase gradient integration is used to construct a phase response for the magnitude STFT. Finally, an inverse STFT is performed to synthesize audio. This model is easy to train when compared to other state-of-the-art methods, allowing for musicians to have full control over the tool. 

This paper presents improvements over the methods outlined in \cite{colonel}. First, this paper incorporates a phase construction method not utilized in \cite{colonel}, which allows for music synthesis through activating the autoencoder's latent space. The method presented in \cite{colonel} requires an input time signal's phase response to construct a time signal at the output.  Second, this work explores asymmetrical autoencoder design via input augmentation, which \cite{colonel} did not. Third, this work compares the performance of several cost functions in training the autoencoder, whereas \cite{colonel} only used mean squared error (MSE).

We have coded an open-source implementation of our method in Python, available at $github.com/JTColonel/canne\_synth$.  

\section{Autoencoding Neural Networks}
\subsection{Mathematical Formulation}
An autoencoder is typically used for unsupervised learning of an encoding scheme for a given input domain, and is comprised of an encoder and a decoder
\cite{stacked}. For our purposes, the encoder is forced to shrink the dimension of an input into a latent space using a discrete number of values, or ``neurons.'' The decoder then expands the dimension of the latent space to that of the input, in a manner that reconstructs the original input. 

We will first restrict our discussion to a single layer model where the encoder maps an input vector $x \in \mathbb{R}^d $ to the hidden layer $y \in \mathbb{R}^e $, where $d>e$. Then, the decoder maps $y$ to $\hat{x} \in \mathbb{R}^d $. In
this formulation, the encoder maps $x \rightarrow y $ via
\begin{equation} \label{Wx:encode}
y = f(Wx+b)
\end{equation}
where $W \in \mathbb{R}^{(e \times d)}$, $b \in \mathbb{R} ^e $, and
$f(\cdotp)$ is an activation function that imposes a non-linearity in
the neural network. The decoder has a similar formulation:
\begin{equation} \label{Wx:decode}
\hat{x} = f(W_{\text{out}} y+b_{\text{out}})
\end{equation}
with $W_{\text{out}} \in \mathbb{R} ^{(d \times e)}$, $b_{out} \in \mathbb{R} ^d $. 

A multi-layer autoencoder acts in much the same way as a single-layer autoencoder. The encoder contains $n>1$ layers and the decoder contains $m>1$ layers. Using equation \ref{Wx:encode} for each mapping, the encoder maps $x \rightarrow x_{1} \rightarrow \ldots  \rightarrow x_{n} $. Treating $x_{n}$ as $y$ in equation \ref{Wx:decode}, the decoder maps $x_{n} \rightarrow x_{n+1} \rightarrow \ldots \rightarrow x_{n+m} = \hat{x}$. 

The autoencoder trains the weights of the $W$'s and $b$'s to minimize some cost function. This cost function should minimize the distance between input and output values.The choice of activation functions $f(\cdotp)$ and cost functions relies on the domain of a given task.

\begin{figure*} [t]
  \centering
\caption{\label{Ann_topology} Autoencoder Topology used. Each layer is fully-connected and feed-forward. The value above each layer denotes the width of the hidden layer.}
\label{recon}
\includegraphics[width=\textwidth]{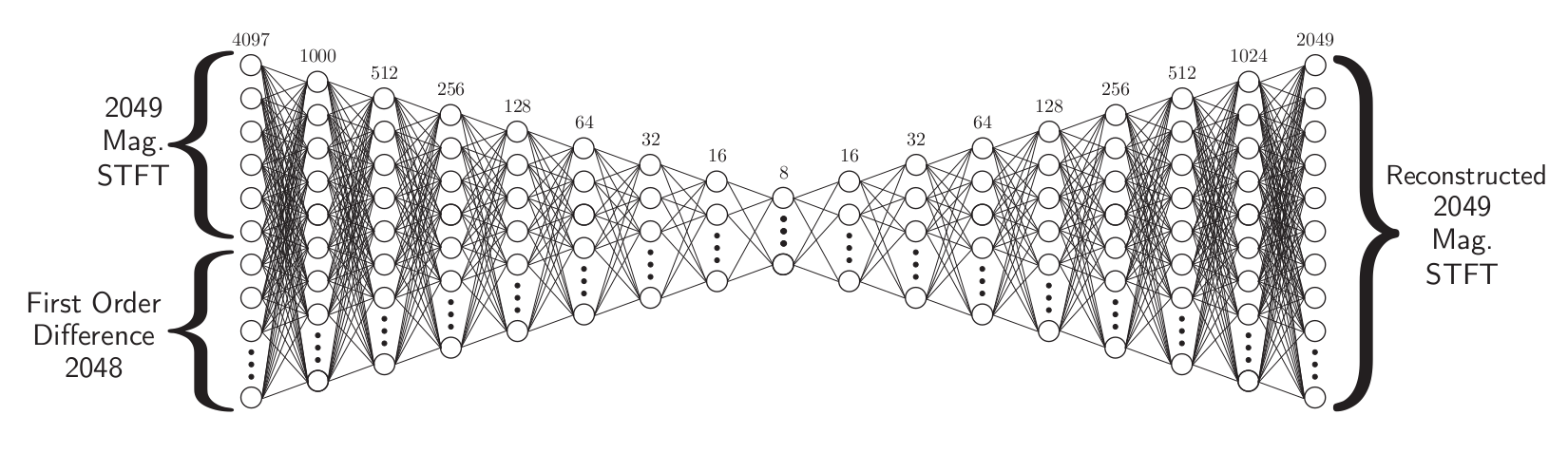}
\end{figure*}

\subsection{Learning Task Description}
In our work we train a multi-layer autoencoder to learn representations of musical audio. Our aim is to train an autoencoder to contain high level, descriptive audio features in a low dimensional latent space that can be reasonably handled by a musician. As in the formulation above, we impose dimension reduction at each layer of the encoder until we reach the desired dimensionality.

The autoencoding neural network used here takes 2049 points from a
4096-point magnitude STFT $s_{n}(m)$ as its target, where $n$ denotes the frame index of the STFT and $m$ denotes the frequency index. Each frame is normalized to $[0,1]$. 

The cost function used in this work is
spectral convergence (SC) \cite{spectral}:
\begin{equation}\label{SC}
C(\theta_n) = \sqrt{\frac{\sum_{m=0}^{M-1}(s_{n}(m)-\hat{s}_{n}(m))^2}{\sum_{m=0}^{M-1}(s_{n}(m))^2}}
\end{equation}
where $\theta_n$ is the autoencoder's trainable weight variables,$s_{n}(m)$ is the original magnitude STFT frame, $\hat{s}_{n}(m)$ is the reconstructed magnitude STFT frame, and $M$ is the total number of frequency bins in the STFT. 

We fully discuss our decision to use SC in section \ref{NNC}.

\begin{figure*}[t] \label{CF-plots}
\centering
\includegraphics[width=0.33\textwidth]{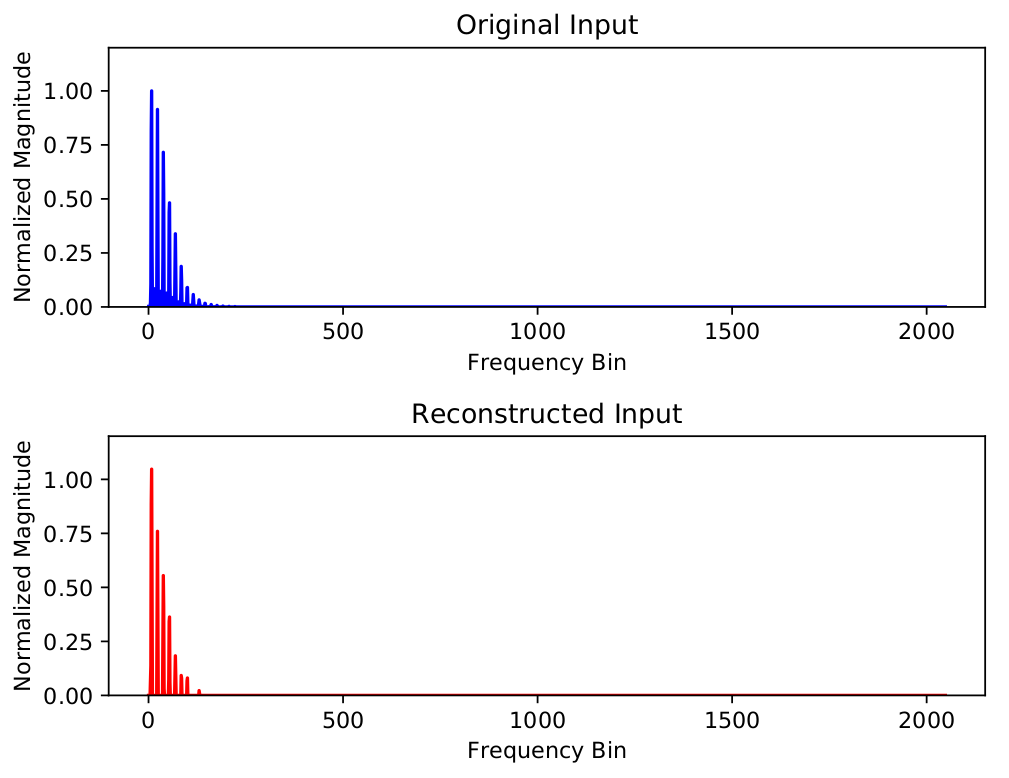} \hfill
\includegraphics[width=0.33\textwidth]{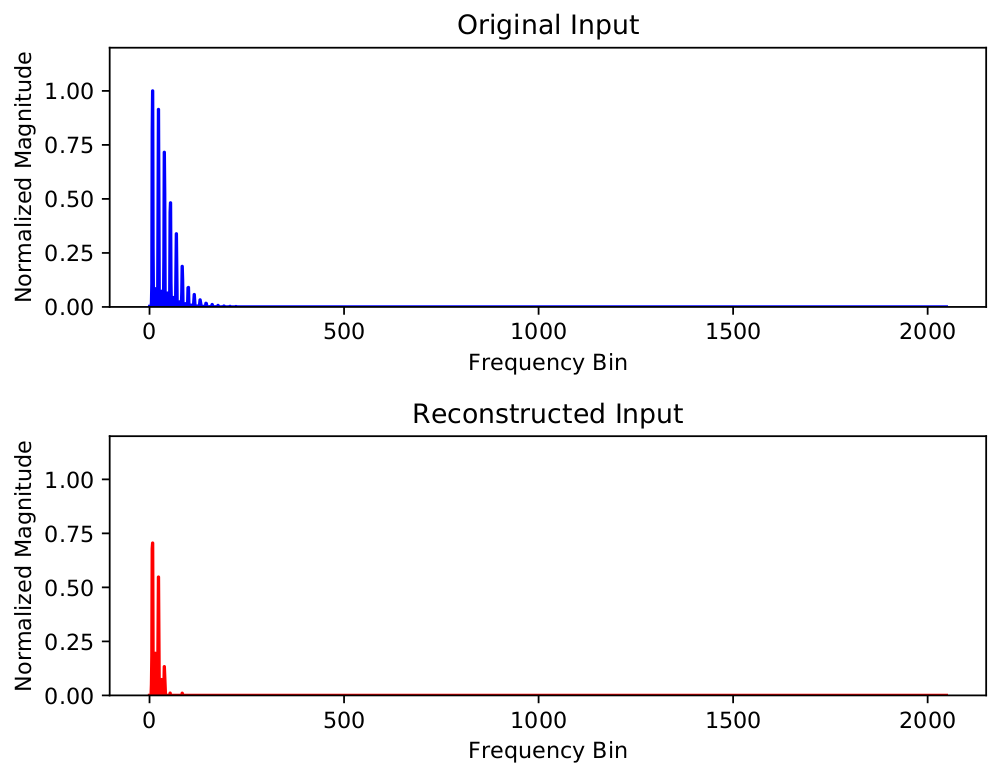} \hfill
\includegraphics[width=0.33\textwidth]{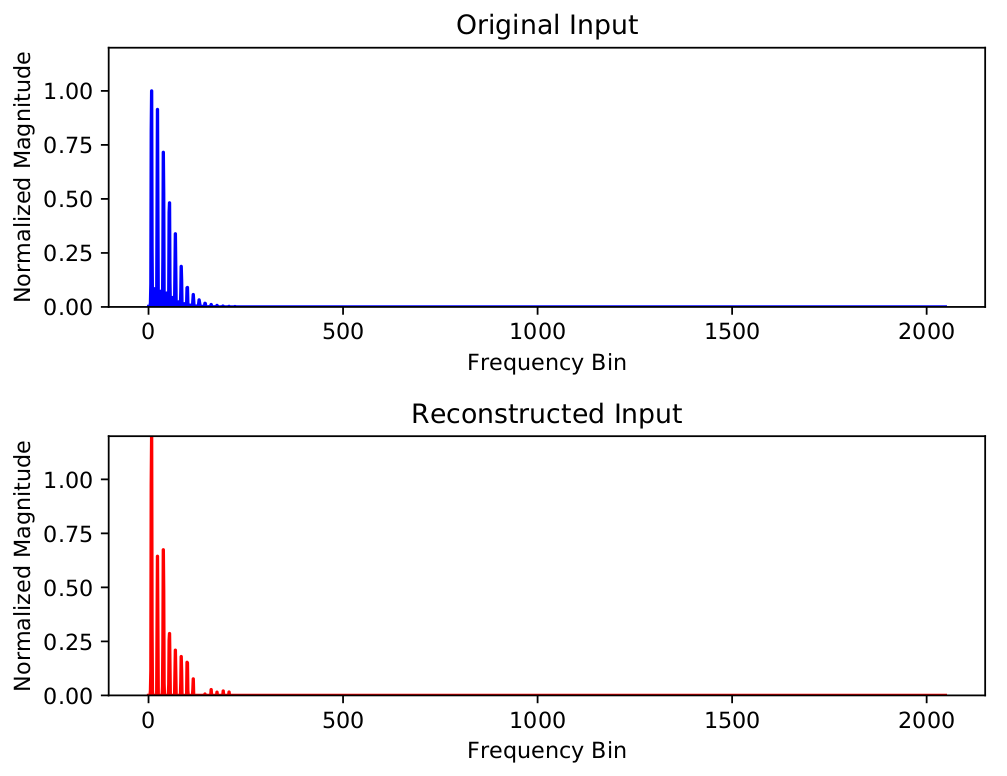}
\caption{\label{CF-Compare}{Sample input and recosntruction using three different cost functions: SC (left), MSE (center), and MAE (right)}} 
\end{figure*}

\subsection{Corpus}
All topologies presented in this paper are trained using approximately $\num[group-separator={,}]{79000}$ magnitude STFT frames, with an additional $\num[group-separator={,}]{6000}$ frames held out for testing and another $\num[group-separator={,}]{6000}$   for validation. This makes the corpus  $\num[group-separator={,}]{91000}$  frames in total. The audio used to generate these frames is composed of five octave C Major scales recorded from a MicroKORG synthesizer/vocoder across 80 patches. 70 patches make up the training set, 5 patches make up the testing set, and 5 patches make up the validation set. These patches ensured that different timbres were present in the corpus. To ensure the integrity of the testing and validation sets, the dataset was split on the ``clip'' level. This means that the frames in each of the three sets were generated from distinct passages in the recording, which prevents duplicate or nearly duplicate frames from appearing across the three sets. 

By restricting the corpus to single notes played on a MicroKORG, the autoencoder needs only to learn higher level features of harmonic synthesizer content. These tones often have time variant timbres and effects, such as echo and overdrive. Thus the autoencoder is also tasked with learning high level representations of these effects. We have made our corpus available as both \textit{.wav} files and as a \textit{.npy} record. Furthermore, we provide a script that creates new corpora, formatted for training our autoencoder, given a \textit{.wav} file.

\section{Neural Network Construction} \label{NNC}
\subsection{Topology}

A fully-connected, feed-forward neural network acts as our autoencoder. Refer to Figure \ref{Ann_topology} for an explicit diagram of the network architecture. Our decisions regarding activation functions, input augmentation, and additive biases are discussed below.  

\subsection{ReLU Activation Function}
In order for training to converge, the rectified linear unit (ReLU) was chosen as the activation function for each layer of the autoencoder \cite{relu}.  The ReLU is formulated
as
\begin{equation}
f(x) = 
\left\{
\begin{array}{ll}
      0 &,  x < 0 \\
      x &,  x \geq 0 \\
\end{array} 
\right. \end{equation}

This activation function has the benefit of having a gradient of either zero or one, thus avoiding the vanishing gradient problem \cite{vangrad}.  

Following \cite{colonel}, we found that using additive bias terms $b$ in Equation \ref{Wx:encode} created a noise floor within the autoencoder, thus we chose to leave them out in the interest of musical applications.  

\subsection{Spectral Convergence Cost Function with L2 Penalty}
As mentioned above SC (Eqn. \ref{SC}) was chosen as the cost function for this autoencoder instead of mean squared error (MSE)
\begin{equation}
C(\theta_n) = \frac{1}{M}\sum_{m=0}^{M-1}(s_{n}(m)-\hat{s}_{n}(m))^2
\end{equation} 
or mean absolute error (MAE)
\begin{equation}
C(\theta_n) = \frac{1}{M}\sum_{m=0}^{M-1}|s_{n}(m)-\hat{s}_{n}(m)|
\end{equation} 
The advantages of using SC as a cost function are twofold. First, its numerator penalizes the autoencoder in much the same way mean squared error (MSE) does. That is to say, reconstructed frames dissimilar from their input are penalized on a sample-by-sample basis, and the squared sum of these deviations dictates magnitude of the cost. 

The second advantage, and the primary reason SC was chosen over MSE, is that its denominator penalizes the autoencoder in proportion to the total spectral power of the input signal. Because the training corpus used here is comprised of ``simple'' harmonic content (i.e. not chords, vocals, percussion, etc.), much of a given input's frequency bins will have zero or close to zero amplitude.  SC's normalizing factor gives the autoencoder less leeway in reconstructing harmonically simple inputs than MSE or MAE. Refer to Figure \ref{CF-Compare} for diagrams demonstrating the reconstructive capabilities each cost function produces.

\begin{figure*}[t] 
\centering
\includegraphics[width=0.45\textwidth]{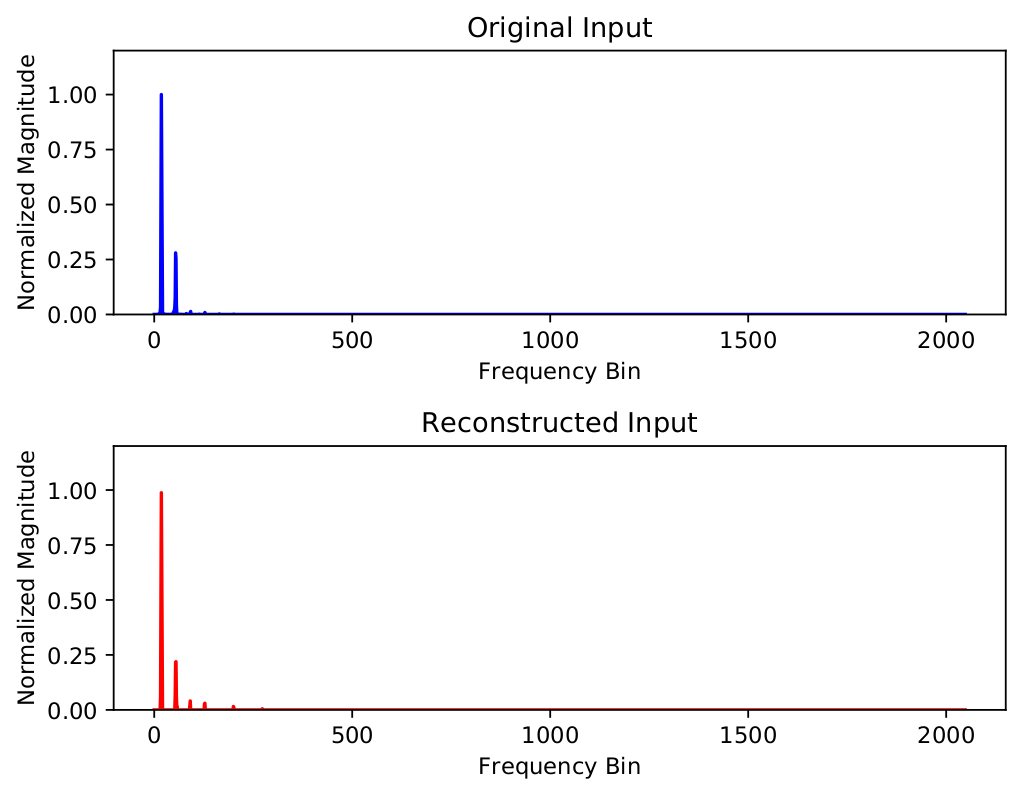} \hfill
\includegraphics[width=0.45\textwidth]{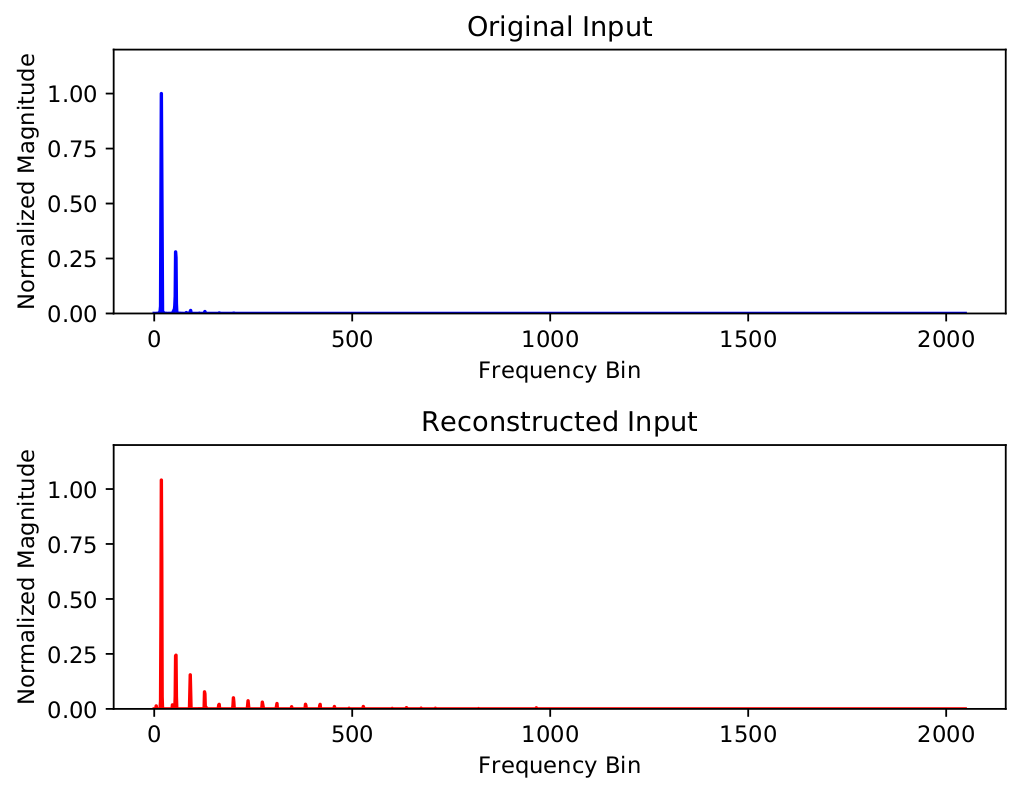}
\caption{\label{mfcc}{Sample input and reconstruction for the first-order appended model (left) and mfcc appended model (right)}} 
\end{figure*}

As mentioned in \cite{colonel}, we found that the autoencoder would not always converge when using SC by itself as the cost function. Thus, we added an L2 penalty to the cost function
\begin{equation}\label{SCL2}
C(\theta_n) = \sqrt{\frac{\sum_{m=0}^{M-1}(s_{n}(m)-\hat{s}_{n}(m))^2}{\sum_{m=0}^{M-1}(s_{n}(m))^2}}+\lambda_{l2} \| \theta_n \|_{_{2}}
\end{equation}
where $\lambda_{l2}$ is a tuneable hyperparameter and
$\| \theta_n \|_{_{2}}$ is the Euclidean norm of the autoencoder's
weights \cite{L2}. This normalization technique encourages the autoencoder to
use smaller weights in training, which we found to improve
convergence. We set  $\lambda_{l2}$ to $10^{-20}$. This value of $\lambda_{l2}$ is large enough to prevent runaway weights while still allowing the SC term to dominate in the loss evaluation.

\subsection{Input Augmentation}
Despite these design choices, we still found the performance of the autoencoder to be subpar. To help the autoencoder enrich its encodings, we augmented its input with higher-order information. We tried augmenting the input with different permutations of the input magnitude spectrum's first-order difference,
\begin{equation}
x_{1}[n] = x[n+1]-x[n]
\end{equation}
second-order difference,
\begin{equation}
x_{2}[n] = x_{1}[n+1]-x_{1}[n]
\end{equation} and Mel-Frequency Cepstral Coefficients (MFCCs). 

MFCCs have seen widespread use in automatic speech recognition, and can be thought of as the "spectrum of the spectrum." In our application, a 512 band mel-scaled log-transform of $s_{n}(m)$ is taken. Then, a 256-point discrete-cosine transform  is performed. The resulting aplitudes of this signal are the MFCCs. Typically the first few cepstral coefficients are orders of magnitude larger than the rest, and we found this to impede training. Thus before appending the MFCCs to our input, we throw out the first five cepstral values and normalize the rest to [-1,1]. 

\subsection{Training Implementation}
All audio processing was
handled by the librosa Python library \cite{librosa}. In this
application, librosa was used to read \textit{.wav} files
sampled at 44.1kHz, perform STFTs of length 4096 with centered
Hann window, hop length 1024 (25\%), and write 16-bit PCM \textit{.wav} files with sampling frequency 44.1kHz from reconstructed magnitude STFT frames.
  
The neural network framework was handled using TensorFlow \cite{tensorflow}. 
All training used the Adam method for stochastic gradient descent with mini-batch size of 200 \cite{ADAM} for 300 epochs. ALl models were trained on an NVIDIA GeForce GTX Titan X GPU. A checkpoint file containing the trained weights of each autoencoder topology was saved once training was finished.

\subsection{Task Performance/Evaluation}
Table \ref{loss} shows the SC loss on the validation set after training. For reference, an autoencoder that estimates all zeros for any given input has a SC loss of $0.843$. 

As demonstrated, the appended inputs to the autoencoder improve over the model with no appendings. Our results show that while autoencoders are capable of constructing high level features from data unsupervised, providing the autoencoder with common-knowledge descriptive features of an input signal can improve its performance. 
\begin{table}[t] 
  \caption{\label{loss} Autoencoder validation set SC loss and Training Time}
  \label{MSE3}
  \centering
  \begin{tabular}{|c|c|c|}
    \hline Input Append     & Validation SC		& Training Time     \\\hline
    No Append	&	$0.257 $  &	25 minutes \\	
    $1^{st}$ Order Diff & $0.217 $ &	51 minutes	 \\
    $2^{nd}$ Order Diff  & $0.245 $ &	46 minutes	 \\
    $1^{st}$ and $2^{nd}$ Order Diff  & $0.242 $ &	69 minutes	 \\
    MFCCs   & $0.236 $ &	52 minutes	 \\\hline
    %Half MFCCs   & $0.238 $ &	46 minutes	 \\
  \end{tabular}
\end{table}

The model trained by augmenting the input with the signal's $1^{st}$ order difference ($1^{st}$-order-appended model) outperformed every other model. Compared to the $1^{st}$-order-appended model, the MFCC trained model often inferred overtonal activity not present in the original signal (Figure \ref{mfcc}). While it performs worse on the task than the $1^{st}$-order-append model, the MFCC trained model presents a different sound palette that is valid for music synthesis. Options for training the model with different appending schemes are available in our implementation.

\section{Audio Synthesis}
\subsection{Spectrogram Generation}
The training scheme outline above forces the autoencoder to construct a latent space contained in $\mathbb{R}^{8}$ that contains representations of synthesizer-based musical audio. Thus a musician can use the autoencoder to generate spectrograms  by removing the encoder and directly activating the 8 neuron hidden layer. However, these spectrograms come with no phase information. Thus to obtain a time signal, phase information must be generated as well.

\subsection{Phase Generation with RTPGHI}
Real-time phase gradient heap integration (RTPGHI) \cite{phase} is used to generate the phase for the spectrogram. While the full theoretical treatment of this algorithm is outside the scope of this paper, we present the following synopsis.

The scaled discrete STFT phase gradient $\nabla \phi = (\phi_{\omega},\phi_{t})$ can be approximated by first finding the phase derivative in the time direction $\tilde{\phi}_{t,n}$ 
\begin{equation}
\tilde{\phi}_{t,n}(m)= \frac{aM}{2\gamma}(s_{log,n}(m+1)-s_{log,n}(m-1))+2\pi am/M
\end{equation}
where $s_{log,n}(m)=log(s_{n}(m))$ and $\tilde{\phi}_{t,n}(0,n)=\tilde{\phi}_{t,n}(M/2,n)=0$. Because a Hann window of length 4098 is used to generate the STFT frames, $\gamma=0.25645 \times 4098^{2}$. Then, the phase derivative in the frequency direction is calculated using a first order difference approximation to estimate the phase $\tilde{\phi}_{n}(m)$ using the following algorithm
\begin{equation}
\tilde{\phi}_{n}(m) \leftarrow \tilde{\phi}_{n-1}(m)+\frac{1}{2}(\tilde{\phi}_{t,n-1}(m)+\tilde{\phi}_{t,n}(m))
\end{equation}

An inverse STFT (ISTFT) is then taken using the generated spectrogram and phase to produce a time signal.

An issue arises when using RTPGHI with this autoencoder architecture. A spectrogram generated from a constant activation of the hidden layer contains constant magnitudes for each frequency value. This leads to the phase gradient not updating properly due to the 0 derivative between frames. To avoid this, uniform random noise drawn from [0.999,1.001] is multiplied to each magnitude value in each frame. By multiplying this noise rather than adding it, we avoid adding spectral power to empty frequency bins and creating a noise floor in the signal. 

\section{Python Implementation}
\begin{figure}[t] 
\centerline{\includegraphics[scale=0.5]{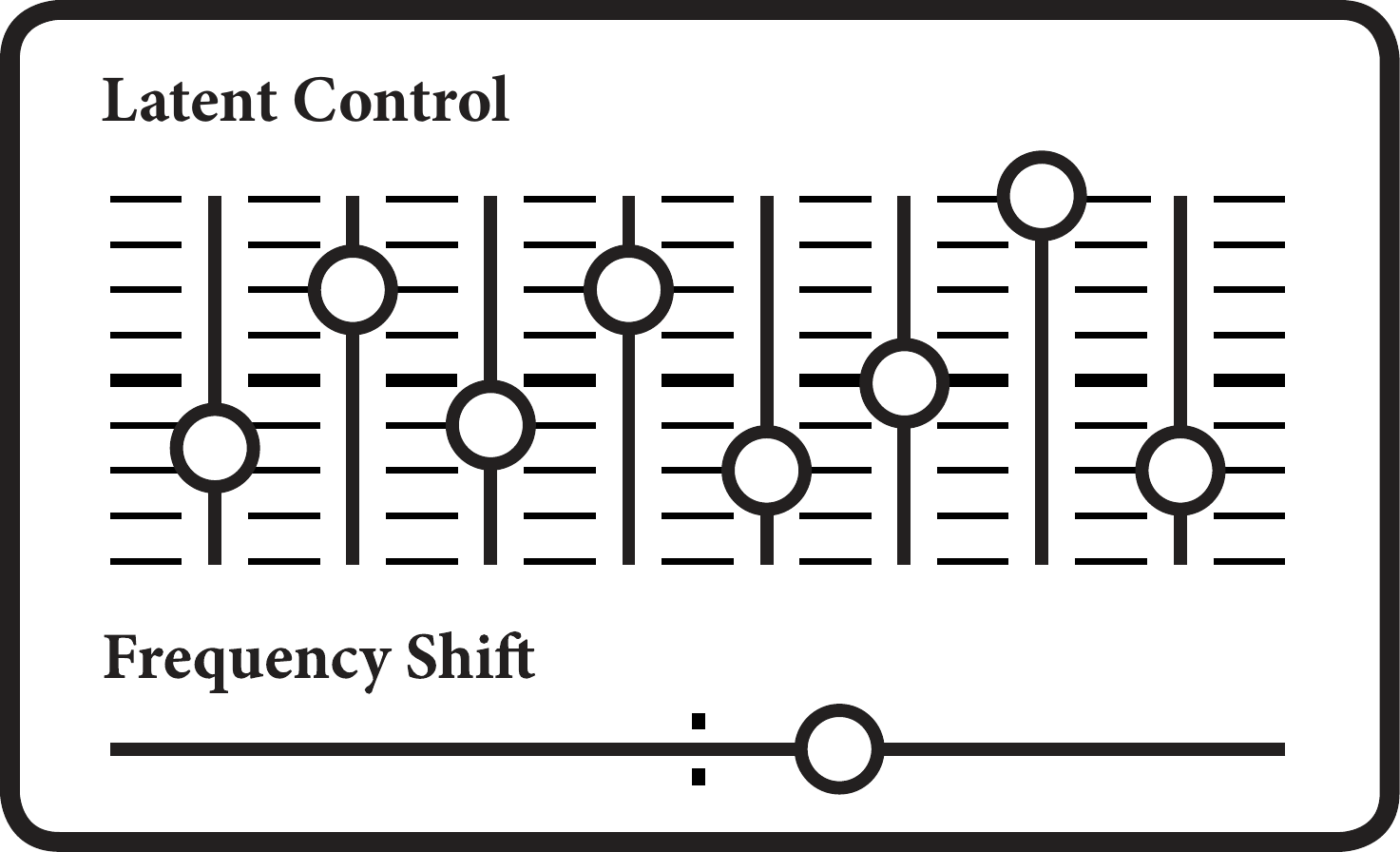}}
\caption{\label{GUI}{Mock-up GUI for CANNe.}} 
\end{figure}
\subsection{CANNe}
We realized a software implementation of our autoencoder synthesizer, ``CANNe (Cooper's Autoencoding Neural Network)'' in Python using TensorFlow, librosa, pygame, soundfile, and Qt 4.0. Tensorflow handles the neural network infrastructure, librosa and soundfile handle audio processing, pygame allows for audio playback in Python, and Qt handles the GUI. 

Figure \ref{GUI} shows a mock-up of the CANNe GUI. A musician controls the eight Latent Control values to generate a tone. The Frequency Shift control performs a circular shift on the generated magnitude spectrum, thus effectively acting as a pitch shift. It is possible, though, for very high frequency content to roll into the lowest frequency values, and vice-versa.

\section{Conclusions}
We present a novel method for musical audio synthesis based on activating the smallest hidden layer of an autoencoding neural network. By training the autoencoder to encode and decode magnitude short-time Fourier transform frames, the autoencoder is forced to learn high-level, descriptive features of audio. Real-time phase gradient heap integration is used to calculate a phase response for the generated magnitude response, thus making an inverse STFT possible and generating a time signal. We have implemented our architecture and algorithm in Python and host the open-source code at $github.com/JTColonel/canne\_synth$.

\section{Acknowledgments}
We would like to thank Benjamin Sterling for helping to code early implementations of the CANNe GUI and Yonatan Katzelnik for helping design and structure our UI. 
%\newpage
\nocite{*}
\bibliographystyle{IEEEbib}
\bibliography{DAFx18_tmpl} % requires file DAFx18_tmpl.bib

\end{document}